\documentclass[twocolumn,floatfix]{aastex631}
\usepackage{hyperref}

\newcommand{\UChicago}{\affiliation{Department of Physics, The University of Chicago, 5640 South Ellis Avenue, Chicago, Illinois 60637, USA}}
\newcommand{\CITA}{\affiliation{Canadian Institute for Theoretical Astrophysics, University of Toronto, 60 St George St,  Toronto, ON M5S 3H8, Canada}}
\newcommand{\UofT}{\affiliation{David A. Dunlap Department of Astronomy and Astrophysics, and Department of Physics, 60 St George St, University of Toronto, Toronto, ON M5S 3H8, Canada}}
\newcommand{\KICP}{\affiliation{Kavli Institute for Cosmological Physics, The University of Chicago, 5640 South Ellis Avenue, Chicago, Illinois 60637, USA}}
\newcommand{\EFI}{\affiliation{Enrico Fermi Institute, The University of Chicago, 933 East 56th Street, Chicago, Illinois 60637, USA}}
\newcommand{\UChicagoAA}{\affiliation{Department of Astronomy and Astrophysics, The University of Chicago, 5640 South Ellis Avenue, Chicago, Illinois 60637, USA}}

\newcommand{\corrections}[1]{{{{#1}}}}

\usepackage{appendix}
\usepackage{bm}

\begin{document}

\title{Gravitational-wave dark siren cosmology systematics from galaxy weighting}

\author[0000-0002-8304-0109]{Alexandra G. Hanselman}
\email{aghanselman@uchicago.edu}
\UChicago
\author[0000-0002-4103-0666]{Aditya Vijaykumar}
\email{aditya@utoronto.ca}
\CITA
\UChicago
\author[0000-0002-1980-5293]{Maya Fishbach}
\CITA
\UofT
\author[0000-0002-0175-5064]{Daniel E. Holz}
\UChicago
\EFI
\UChicagoAA
\KICP

\begin{abstract}
The detection of GW170817 and the measurement of its redshift from the associated electromagnetic counterpart provided the first gravitational wave determination of the Hubble constant ($H_0$), demonstrating the potential power of standard-siren cosmology. In contrast to this bright siren approach, the dark siren approach can be utilized for gravitational-wave sources in the absence of an electromagnetic counterpart: one considers all galaxies contained within the localization volume as potential hosts. When statistically averaging over the potential host galaxies, weighting them by physically-motivated properties (e.g., tracing star formation or stellar mass) could improve convergence. Using mock galaxy catalogs, we explore the impact of these weightings on the measurement of $H_0$. We find that incorrect weighting schemes can lead to significant biases due to two effects: the assumption of an incorrect galaxy redshift distribution, and preferentially weighting incorrect host galaxies during the inference. The magnitudes of these biases are influenced by the number of galaxies along each line of sight, the measurement uncertainty in the gravitational-wave luminosity distance, and correlations in the parameter space of galaxies. We show that the bias may be overcome from improved localization constraints in future GW detectors, a strategic choice of priors or weighting prescription, and by restricting the analysis to a subset of high-SNR events. We propose the use of hierarchical inference as a diagnostic of incorrectly-weighted prescriptions. Such approaches can simultaneously infer the correct weighting scheme and the values of the cosmological parameters, thereby mitigating the bias in dark siren cosmology due to incorrect host-galaxy weighting.

\end{abstract}

\keywords{}
\section{Introduction} \label{sec:intro} 

Measuring the expansion rate of the universe has been a key goal of observational cosmology for almost a century. Specifically, the local expansion rate of the universe, the Hubble constant $H_0$, has recently been a topic of intense debate. Precision measurements of $H_0$ from low redshift probes (e.g. supernovae;~\citealt{2022ApJ...938..113S,2023arXiv230801875U}) and high redshift probes (e.g. cosmic microwave background;~\citealt{2020A&A...641A...6P}) disagree,  giving rise to the ``$H_0$ tension'' (see \citealt{2021CQGra..38o3001D,2023JCAP...11..050F}). Alternate observational probes of $H_0$ are of particular utility in distinguishing whether the tension is due to unmodelled systematics in current observations or physics beyond the standard model of cosmology.

Observations of gravitational waves (GWs) from compact binary coalescences (CBCs) have been proposed as probes of cosmic expansion~\citep{1986Natur.323..310S, 2005ApJ...629...15H}. The luminosity distance to a CBC can be estimated directly from the gravitational waveform~\citep{1986Natur.323..310S, 1993PhRvD..47.2198F, 1994PhRvD..49.2658C} without reference to a distance ladder.  
If the cosmological redshift of the CBC can be estimated by some other means, it is possible to infer the values of $H_0$ and other cosmological parameters governing the expansion history of the universe. For instance, binary neutron stars (BNS) have electromagnetic counterparts which can enable the localization of the source to its host galaxy, yielding a redshift measurement~\citep{2005ApJ...629...15H, 2006PhRvD..74f3006D, 2010ApJ...725..496N, 2013arXiv1307.2638N, 2018Natur.562..545C}. The observation of an electromagnetic counterpart from GW170817 led to the identification of NGC~4993 as the source's host galaxy~\citep{2017PhRvL.119p1101A, 2017ApJ...848L..13A, 2017ApJ...848L..12A,2017Sci...358.1556C,2017ApJ...848L..16S}, yielding a $\sim 15 \%$ measurement of $H_0$ solely from this source~\citep{2017Natur.551...85A}. However, subsequent observing runs of the LIGO-Virgo-KAGRA (LVK) collaboration have to date failed to yield additional mergers with electromagnetic counterparts~\citep{2023PhRvX..13d1039A}.

In the absence of electromagnetic counterparts, Bernard Schutz proposed an alternative ``dark siren'' method where one considers all galaxies in the localization volume of a given CBC as potential hosts~\citep{1986Natur.323..310S,2008PhRvD..77d3512M,2012PhRvD..86d3011D,2018Natur.562..545C,2019ApJ...871L..13F, 2020PhRvD.101l2001G,2019ApJ...876L...7S, 2020ApJ...900L..33P, 2021ApJ...909..218A, 2021JCAP...08..026F,2023ApJ...949...76A,2022arXiv220309238M,2023ApJ...943...56P}. For a typical CBC, the number of galaxies in a typical localization volume is large \citep[e.g., $\sim 408$ for GW170817;][]{2019ApJ...871L..13F}; consequently, the $H_0$ measurement from a single event using this method is generally highly uncertain. However, the expectation is that $H_0$ measurements stacked over multiple events would reduce this uncertainty, enabling an $H_0$ measurement of a few percent. This statistical dark siren technique yields a $\sim 20\%$ measurement of $H_0$ from dark sirens alone in current data~\citep{2023ApJ...949...76A, 2023ApJ...943...56P, 2022arXiv220309238M}. Other methods of redshift identification have been proposed, including using the large-scale two-point cross-correlation between galaxies and GW mergers~\citep{2016PhRvL.116l1302N, 2020ApJ...902...79B,2021PhRvD.103d3520M,2022MNRAS.511.2782C}, identification of features in the mass spectrum~\citep[``spectral sirens'';][]{2012PhRvD..85b3535T,2019ApJ...883L..42F,2021PhRvD.104f2009M,2022PhRvL.129f1102E}, and harnessing information from the equation of state of dense matter~\citep[``Love sirens'';][]{2012PhRvL.108i1101M,2021PhRvD.104h3528C}.

One of the ingredients that goes into the statistical dark siren method is the probability that a particular galaxy is the host of the CBC based on its physical properties~\citep{2012PhRvL.108i1101M, 2019ApJ...871L..13F, 2020PhRvD.101l2001G}. For instance, depending on the delay-time distribution of CBCs, they could either preferentially merge in star-forming galaxies (short delay times) or massive galaxies (long delay times) (see e.g.~\citealt{Adhikari:2020wpn}; \citealt{Vijaykumar:2023bgs})\footnote{See also other works that forward-model the distribution of GW host galaxies based on population synthesis models~\citep{OShaughnessy:2009szr, Lamberts:2016txh, Mapelli:2019bnp, Toffano:2019ekp, Artale:2019doq, Santoliquido:2022kyu,Srinivasan:2023vaa, 2023MNRAS.523.5719R}.}. While analyzing events to infer $H_0$, this information can be folded in by weighting each candidate host galaxy by its luminosity within a certain bandpass that best tracks the desired physical quantities e.g. a galaxy's B-band luminosity as a proxy for star formation rate or K-band luminosity for its total stellar mass~\citep{2003ApJS..149..289B, 2016ApJ...829L..15S}. Note that choosing to not preferentially weight galaxies based on their physical properties amounts to applying equal weights to all galaxies, and is also an imposition of prior belief about the galaxies that host GWs.
Unfortunately, conclusively inferring the host galaxy distribution from data is difficult, owing to poor sky localization of GW sources. Therefore, diagnosing the impact of incorrect weighting schemes is imperative to understanding any systematics associated with the statistical dark siren approach.  

\cite{2023PDU....4001208T} argue that any results obtained using the dark siren approach would be biased in general. However, ~\cite{Gairetal2023} demonstrate explicitly that their arguments are incorrect. In particular, as long as the dark siren cosmology method is applied consistently, the results are unbiased. However, neither of these works  consider potential biases from the weighting of host galaxies. In this work we explore the impact that an incorrect weighting scheme would have on $H_0$ inference. We do so by building mock catalogs of GW sources and their host galaxies under physically-motivated weights, and explore how the inference is affected by changing the weighting schemes. In general, we find that assuming the correct galaxy host weighting scheme leads to an unbiased $H_0$ estimate, but assuming an incorrect weighting scheme can lead to substantial biases. We also note here that these systematics are different compared to other systematics that impact dark siren cosmology, e.g. models for the mass distribution of CBCs~\citep{2024PhRvD.109h3504P} and photometric redshift uncertainties in galaxy surveys~\citep{2023MNRAS.526.6224T}. During the final stages of this work, \cite{Perna2024} completed a related investigation using the MICECAT mock galaxy catalog~\citep{2015MNRAS.448.2987F,2015MNRAS.453.1513C,2015MNRAS.447.1319F,2015MNRAS.447..646C,2015MNRAS.447.1724H}, considering the case where the true GW merger rates follow the galaxy star formation rate with host galaxies weighted by K-band luminosities. When investigating biases due to mismatch between the GW merger rates and recovered weighting schemes, \cite{Perna2024} find broadly consistent conclusions to what we report below. Here, we expand on the different combinations of injection and recovery weighting schemes, investigate the underlying causes for the biases, and propose potential methods to diagnose and mitigate these biases.

The rest of this paper is organized as follows: in \S\ref{sec:methods}, we describe our inference and data generation prescription. In \S\ref{sec:results}, we identify areas for potential bias. We investigate possible diagnostics and discuss various factors that influence potential biases in \S\ref{sec:implications}, and finally summarize our results in \S\ref{sec:discussion}.

\section{Methods} \label{sec:methods}
\subsection{Inferring $H_0$ using a Bayesian scheme} \label{sec:math}

We use the 
$H_0$ inference prescription outlined in \cite{Gairetal2023} with some modifications that we summarize below. Let us assume we have a set of $N_{\rm GW}$ gravitational wave observations with observed data, $\{\hat{x}\}$, where we take the only important quantity to be the observed luminosity distances $\{\hat{d_L}\}$ to be consistent with \cite{Gairetal2023}\footnote{In more realistic analyses, other GW parameters, such as sky probability, are relevant. Since we are only considering this toy model, we do not write out the full likelihood here, but point readers to~\cite{Gairetal2023} and references therein for a full derivation.}. Using Bayes' theorem, the posterior on $H_0$ is given by
\begin{equation}
p(H_0 | \{\hat{d_L}\} ) \propto \mathcal{L}( \{\hat{d_L}\} | H_0 ) p(H_0) \; ,
\end{equation}
where $p(H_0)$ is the prior on $H_0$, which we take to be uniform, and $\mathcal{L}( \{\hat{d_L}\} | H_0 )$ is the likelihood of observing the data $\{\hat{d_L}\} $ given a value for $H_0$. The likelihood can be further written as \citep{Gairetal2023}:
\begin{equation}
\mathcal{L}( \hat{d_L}^i | H_0 ) = \frac{\int dz\, \mathcal{L}_{\rm GW}( \hat{d_L}^i | d_L(z,H_0) ) \, p_{\rm CBC}(z,w)}{\int dz \, P_{\rm det}^{\rm GW}(\hat{d_L}^i | d_L(z,H_0)) \, p_{\rm CBC}(z,w)} \; ,
\end{equation}
where $\mathcal{L}_{\rm GW}$ is the likelihood of 
measuring an observed luminosity distance 
$\hat{d_L}^i$ given a galaxy with true luminosity distance $d_L(z,H_0)$, and $P_{\rm det}^{\rm GW}$ is the GW detection probability, where we take a GW to be detected if its measured luminosity distance is positive and less than the detection threshold, which we define as $\hat{d_L}^{\rm thr}=1550\,\textrm{Mpc}$ to be consistent with \cite{Gairetal2023}, unless otherwise specified. We also include an extra term $w$ in the likelihood, which allows us to assign weights to galaxies in our catalog (see e.g. \citealt{2019ApJ...871L..13F,2020PhRvD.101l2001G} and references therein). In this work, we take weights that are conditionally independent of redshift and that account for physically-motivated quantities, or use equal weights to consider the case where no preferential probability is given to any galaxy. We provide further details on these weights in \S\ref{sec:UMcat}. We assume the likelihood $\mathcal{L}_{\rm GW}$ is a Gaussian such that
\begin{widetext}
\begin{equation}
 \mathcal{L}_{\rm GW}( \hat{d_L}^i | d_L(z,H_0) ) = \frac{1}{\sqrt{2\pi}A d_L(z,H_0)} \exp\left[-\frac{1}{2} \frac{(\hat{d_L}^i - d_L(z,H_0))^2}{(A d_L(z,H_0))^2}\right] \; ,
 \label{eq:LGW}
\end{equation}
where $A$ is a constant fractional error. This yields a detection probability of
\begin{eqnarray}
    P_{\rm det}^{\rm GW}(\hat{d_L} | d_L(z_j^{\rm gal},H_0)) & = & \int_{-\infty}^{\infty} \Theta(\hat{d_L}^{\rm thr} - \hat{d_L}) \Theta(\hat{d_L}) \mathcal{L}_{\rm GW} (\hat{d_L} | d_L(z_j^{\rm gal},H_0)) d \hat{d_L} \nonumber \\
    & = & \frac{1}{2} \left[\textrm{erf} \left( \frac{1}{\sqrt{2} A}\right) - \textrm{erf} \left( \frac{d_L(z_j^{\rm gal},H_0) - \hat{d_L}^{\rm thr}}{\sqrt{2} A d_L(z_j^{\rm gal},H_0)}\right) \right] \; ,
    \label{eq:pdet}
\end{eqnarray}
\end{widetext}
 where we add an extra bound on $\hat{d_L}$ to ensure that the measured luminosity distance is always positive. Finally, $p_{\rm CBC} (z,w) = \sum_{j=1}^{N_{\rm GAL}} w_j \delta (z - z_j^{\rm gal})$ is the probability that a CBC is at a given redshift $z$, which we assume to be a delta function for each galaxy in our catalog, as we take galaxy redshifts to be perfectly known (see \S\ref{sec:UMcat}), and is weighted by the probability of any galaxy to be the host. In the above prescription, we forgo the $(1+z)^{-1}$ conversion from source to detector frame rate; this will not affect our results since they are consistent in both the simulated dataset and the recovery procedure. These assumptions simplifies the likelihood to: 
\begin{equation}
\mathcal{L}( \hat{d_L}^i | H_0 ) = \frac{ \sum_{j=1}^{N_{\rm GAL}} \mathcal{L}_{\rm GW}( \hat{d_L}^i | d_L(z_j^{\rm gal},H_0) ) w_j }{ \sum_{j=1}^{N_{\rm GAL}} P_{\rm det}^{\rm GW}(\hat{d_L}^i | d_L(z_j^{\rm gal},H_0)) w_j } \; ,
\end{equation}
with $\mathcal{L}_{\rm GW}$ and $P_{\rm det}^{\rm GW}$ now given by Eq.~\ref{eq:LGW} and Eq.~\ref{eq:pdet}. Note that in the above prescription, we ignore sky positions and instead vary the total number of galaxies in a given line of sight
, $N_{\rm GAL}$, as a proxy for varying sky position uncertainty. For reference, an $N_{\rm GAL}$ of 10,000 assuming a galaxy number density of 0.01 $\mathrm{Mpc}^{-3}$ would yeld a localization volume of approximately $10^6 \, \mathrm{Mpc}^{3}$.

\subsection{Mock galaxy catalog} \label{sec:UMcat}

The mock catalog we use in this work is created following the \textsc{UniverseMachine} semi-analytical galaxy formation simulations \citep{Behroozietal2019}. \textsc{UniverseMachine} starts off with a pure dark matter simulation and populates galaxies into halos using a Monte Carlo scheme while ensuring that their derived properties (star formation histories, stellar masses, etc.) are consistent with a wide range of observations. For our purposes, \textsc{UniverseMachine} provides a distribution of galaxies with physical properties such as stellar mass (SM) and star formation rate (SFR) for bins in redshift\footnote{We use the publicly available \textsc{UniverseMachine} dataset from \href{https://halos.as.arizona.edu/UniverseMachine/DR1/SFR/}{https://halos.as.arizona.edu/UniverseMachine/DR1/SFR/}, created using the Bolshoi-Planck~\citep{2011ApJ...740..102K} dark matter simulation box with side length $250\,h^{-1}\,{\rm Mpc}$ and $2048^3$ particles.}. In this work, we take our physically-motivated weights to be either the galaxy stellar mass or star formation rate, e.g. $w_i \propto \{\mathrm{SM}_i, \mathrm{SFR}_i\}$. These are common choices, although in more realistic analyses, the typically chosen weights are a galaxy's K-band or B-band luminosity as a proxy for either SM or SFR, respectively (see e.g.~\cite{2019ApJ...871L..13F,2023ApJ...949...76A}). 
\begin{figure}[tbhp]
    \centering
    \includegraphics[width =\linewidth]{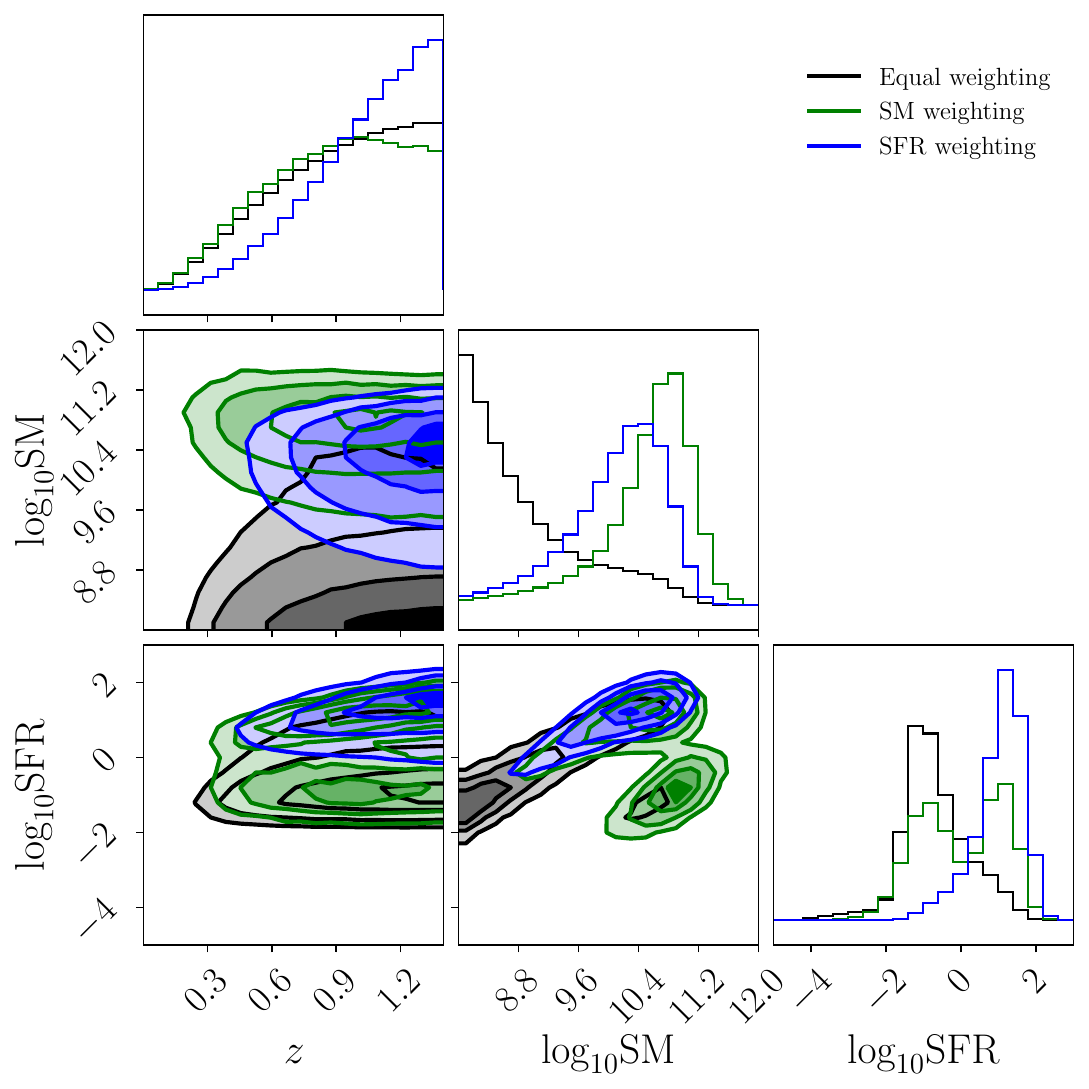}
    \caption{\textsc{UniverseMachine} mock catalog distribution. Notice that the weighted redshift distributions are different---the redshift evolution of the number density follows $\sim(1+z)^{2.5}$ when weighting  with SFR and $\sim(1+z)^{-0.65}$ when weighting with SM \citep{Vijaykumar:2023bgs}. We also note that SFR and SM are correlated, although the correlations change depending on which weighting scheme is used.}
    \label{fig:UMLCdists}
\end{figure}
Throughout this paper we assume that \textsc{UniverseMachine} provides a complete catalog with galaxy redshifts, stellar masses, and star formation rates that are perfectly known\footnote{In reality, these quantities would have observational uncertainties, e.g. photometric redshift measurements have a typical fractional uncertainty of a few percent. These uncertainties would need to be accounted for in the likelihood. Mismodeling the shape of the uncertainty regions could also impact the measurement of  $H_0$~\citep[see][for a discussion on mismodelling redshift uncertainties]{2023MNRAS.526.6224T}.}. We draw a million galaxies from a uniform in comoving volume distribution for redshifts less than 1.4 and bin these draws into the redshift bins provided by \textsc{UniverseMachine}. 
For each galaxy we draw stellar mass (down to $10^8 M_\odot$) and star formation rate values from distributions contained in each redshift bin.
The mock catalog thus created contains physical distributions (i.e. mass, SFR, redshift distributions) consistent with \textsc{UniverseMachine}, but does not include effects of galaxy clustering. We have also tested the analysis below using the second version of the MICECAT mock galaxy catalog~\citep{2015MNRAS.448.2987F,2015MNRAS.453.1513C,2015MNRAS.447.1319F,2015MNRAS.447..646C,2015MNRAS.447.1724H,2022PhRvD.106l3510H}, which includes galaxy clustering as well as SM and SFR values, and find similar results to what we report below (see Appendix~\ref{sec:MICECAT}). \cite{Perna2024} find similar conclusions when investigating potential biases using the first version of the MICECAT catalog using galaxy luminosities.

\begin{figure*}[!tbh]
    \centering
    \includegraphics[width = \linewidth]{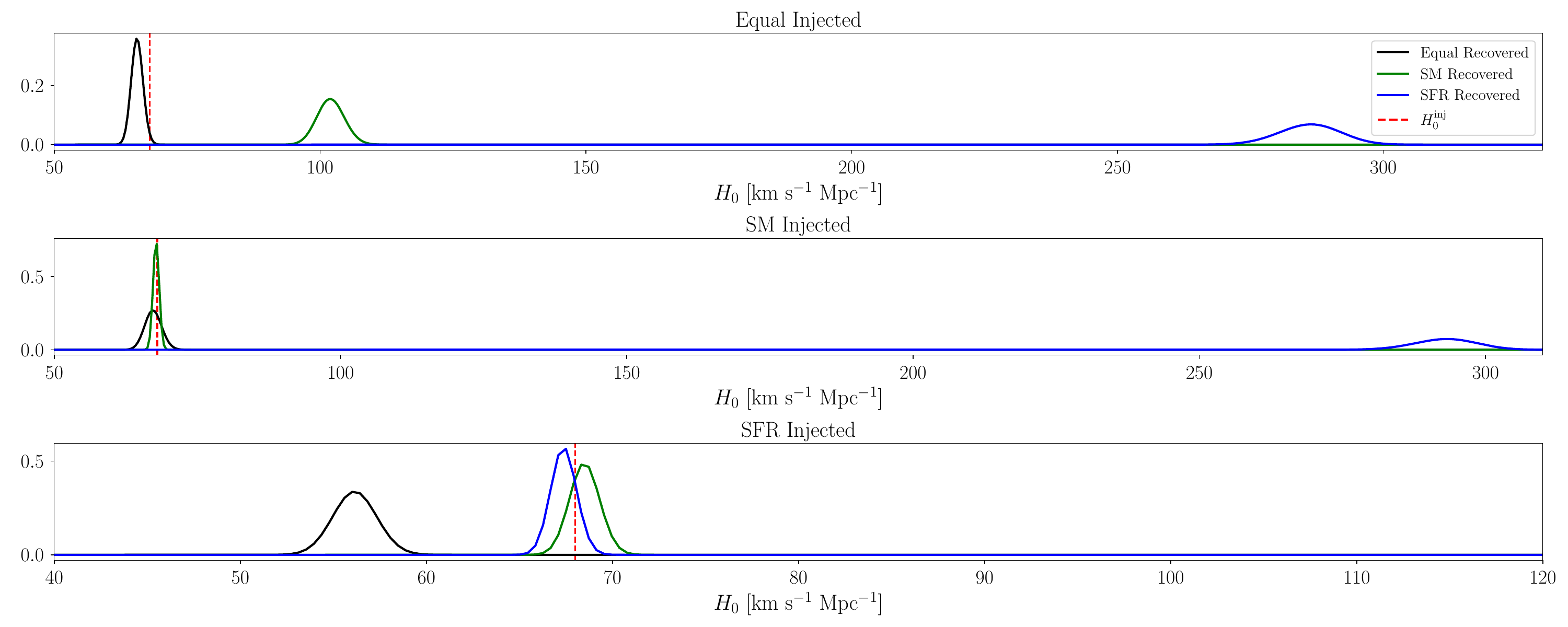}
    \caption{Inference in $H_0$ using the \textsc{UniverseMachine} mock catalog from 20,000 GW events coming from lines of sight each containing 10,000 galaxies. Each GW event has a $20\%$ error ($A=0.2$) in luminosity distance. Assuming the correct weighting scheme leads to an unbiased $H_0$ recovery, but assuming an incorrect weighting scheme can lead to extremely large biases.
    }
    \label{fig:Bias-single-param}
\end{figure*}

The distributions for our mock catalog are shown in Fig.~\ref{fig:UMLCdists}. We see that weighting by SM (green) or SFR (blue) leads to different redshift distributions. As mentioned in \cite{Vijaykumar:2023bgs}, if host galaxies are weighted solely by their star formation rates, the redshift evolution of the galaxy number density is roughly proportional to $\sim(1+z)^{2.5}$, whereas solely weighting by their total stellar masses would be roughly proportional to $\sim(1+z)^{-0.65}$ in the range of redshifts we consider. We also see from the two-dimensional distribution in Fig.~\ref{fig:UMLCdists} that SM and SFR are positively correlated with each other. However, due to the two galaxy branches corresponding to star-forming and quiescent galaxies (see e.g. \cite{Vijaykumar:2023bgs}), although weighting with SFR conserves the positive correlation with SM, weighting with SM leads to a split of hosts between the two branches. We will comment more on this asymmetry in \S\ref{sec:results}.

In our simulations, we first randomly assign each galaxy in our mock catalog to a line of sight, with each line of sight containing $N_{\rm GAL}$ galaxies.
Doing this effectively gives us a constant number density of galaxies (or, alternatively, a constant sky area) along each line of sight. We then generate GW events drawn from our mock catalog weighted by a true (correct) injected weighting scheme.
We then calculate an observed luminosity distance for each GW event by scattering the true luminosity distance of the host galaxy by a normal distribution (see Eq.~\ref{eq:LGW}). If an observed luminosity distance is positive and less than the threshold luminosity distance, $\hat{d_L}^{\rm thr}$, we say that GW event is detected and take the first $N_{\rm GW}$ detected GW events to use in our inference. We take only the lines of sight that contain GW events and calculate
a posterior on $H_0$ using the method described in \S\ref{sec:math}, now assuming a different `recovery’ weighting scheme. Note that this is slightly different from the procedure in \cite{Gairetal2023}, where multiple GWs were assumed to all be coming from a single line of sight. This difference more realistically encapsulates that GWs typically come from different lines of sight\footnote{Note that when galaxies also have measurement uncertainties in redshift, \cite{Gairetal2023} demonstrate that having a similar number of events to galaxies along a single line of sight leads to a bias. Our toy model does not consider galaxy redshift uncertainties, and thus this potential bias is not relevant to our results. Nonetheless, our prescription will not be impacted even if galaxy redshift uncertainties are considered as we always ensure each line of sight contains many more galaxies than GW events.}.

\section{Identifying potential biases} \label{sec:results}

We follow the prescription outlined in Section~\ref{sec:math} using three injection sets where (i) all galaxies are equally likely to be hosts, (ii) mergers follow the galaxy stellar mass, and (iii) mergers follow the galaxy star formation rate. A very conservative uniform $H_0$ prior of $H_0 = [40 ,450] \, {\rm km\, s^{-1} \, Mpc^{-1}}$ is used. We use an injected $H_0$ value of  $H_0^{\rm inj}=68 \,{\rm km\, s^{-1} \, Mpc^{-1}}$ to be consistent with the initial parameters of \textsc{UniverseMachine}. Even for the highest value in the prior, the hard edge in our mock catalog at $z=1.4$ is above the luminosity distance threshold considered. For each injection set, we recover $H_0$ using the three weighting schemes described above. 
\begin{figure*}[htb]
    \centering
    \includegraphics[width = \linewidth]{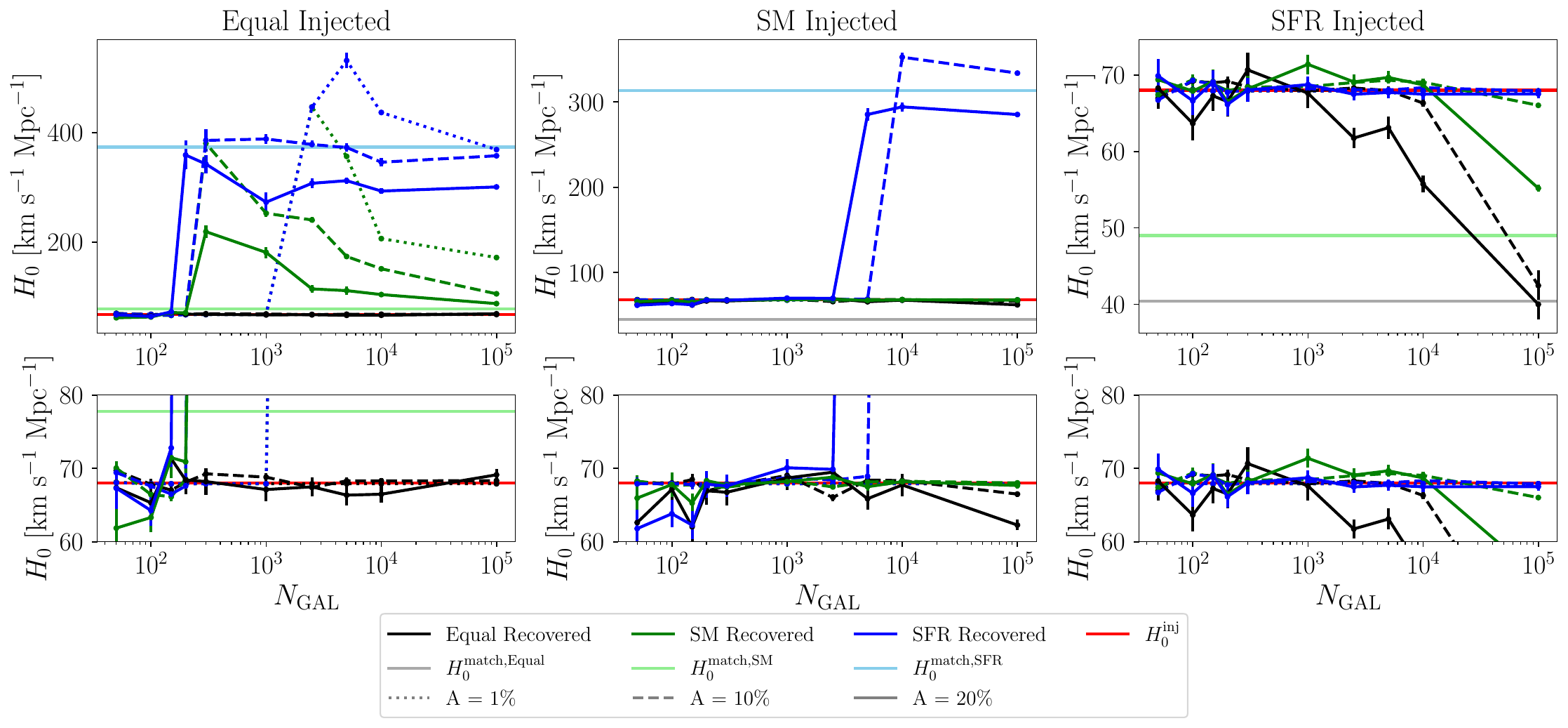}
    \caption{{$H_0$ recovery as a function of $N_{\rm GAL}$ for three choices of fractional error, $A$, for all injection-recovery weighting schemes. The errorbars correspond to the $1\sigma$ uncertainty in the posteriors when observing $2 \times N_{\rm GAL}$ GW events.} We see that, for small enough $N_{\rm GAL}$, the $H_0$ inference remains unbiased for all recovery weighting schemes, and for large enough $N_{\rm GAL}$, the biases asymptote to the theoretical best {`matched'} $H_0$ values (horizontal lines) from matching up the GW luminosity distance distribution with the weighted galaxy redshift distribution {(see the titles in Fig.~\ref{fig:dist-matching})}. These regimes correspond to the ``well-localized" and ``uninformative" regimes described in the main text. Decreasing the fractional error shifts the end of the ``well-localized" regime and beginning of the ``uninformative" regime to higher $N_{\rm GAL}$.
    }
    \label{fig:Bias-per-pcerr}
\end{figure*}

An example posterior on $H_0$ for all injection-recovery schemes is shown in Fig.~\ref{fig:Bias-single-param}. In these results, we {draw} 20,000 {GW events contained within lines of sight, where each line of sight contains} 10,000 galaxies. {Each GW event has a} $20\%$ error {(corresponding to $A=0.2$ in Eq.~\ref{eq:LGW})} in luminosity distance. We use 20,000 GW detections to obtain a tight convergence on the final $H_0$ posterior. Using a smaller subset of detections preserves the magnitude of the bias {(albeit with larger statistical uncertainty)}, but simply adds scatter around the final $H_0$ posterior obtained using the 20,000 detections. Fig.~\ref{fig:Bias-single-param} shows that if we assume the correct weighting scheme, we remain unbiased in our estimate of $H_0$. However, if we assume the incorrect weighting scheme, very large biases become apparent. One striking result is that, when the true distribution follows the galaxy star formation rate, recovering with equal weights results in a bias. We also see that if the true distribution follows SFR, weighting with SM remains unbiased for these initial conditions. However, if the true distribution of mergers follows SM, weighting with SFR is extremely biased. This is due to the asymmetry in the weighted SM--SFR correlations in the \textsc{UniverseMachine} catalog we discussed in \S\ref{sec:UMcat}. As we see in two-dimensional SM--SFR distribution in Fig.~\ref{fig:UMLCdists}, weighting with SFR maintains the positive correlation between SM and SFR in the star-forming branch, which leads to an unbiased estimate of $H_0$ if we were to incorrectly assume the galaxies follow a SM-weighted distribution due to this positive correlation. On the other hand, if the true host probabilities follow SM, we see in Fig.~\ref{fig:UMLCdists} that the most likely galaxy hosts now split between the star-forming and quiescent branches, disrupting the purely positive correlation between SM and SFR, and leading to a bias in $H_0$ when incorrectly assuming the hosts follow SFR. We discuss the effects behind the biases seen in Fig.~\ref{fig:Bias-single-param} in more depth in the following subsections, as well as in Appendix~\ref{sec:mockcats}.

\begin{figure*}
\centering
    \includegraphics[width = \linewidth]{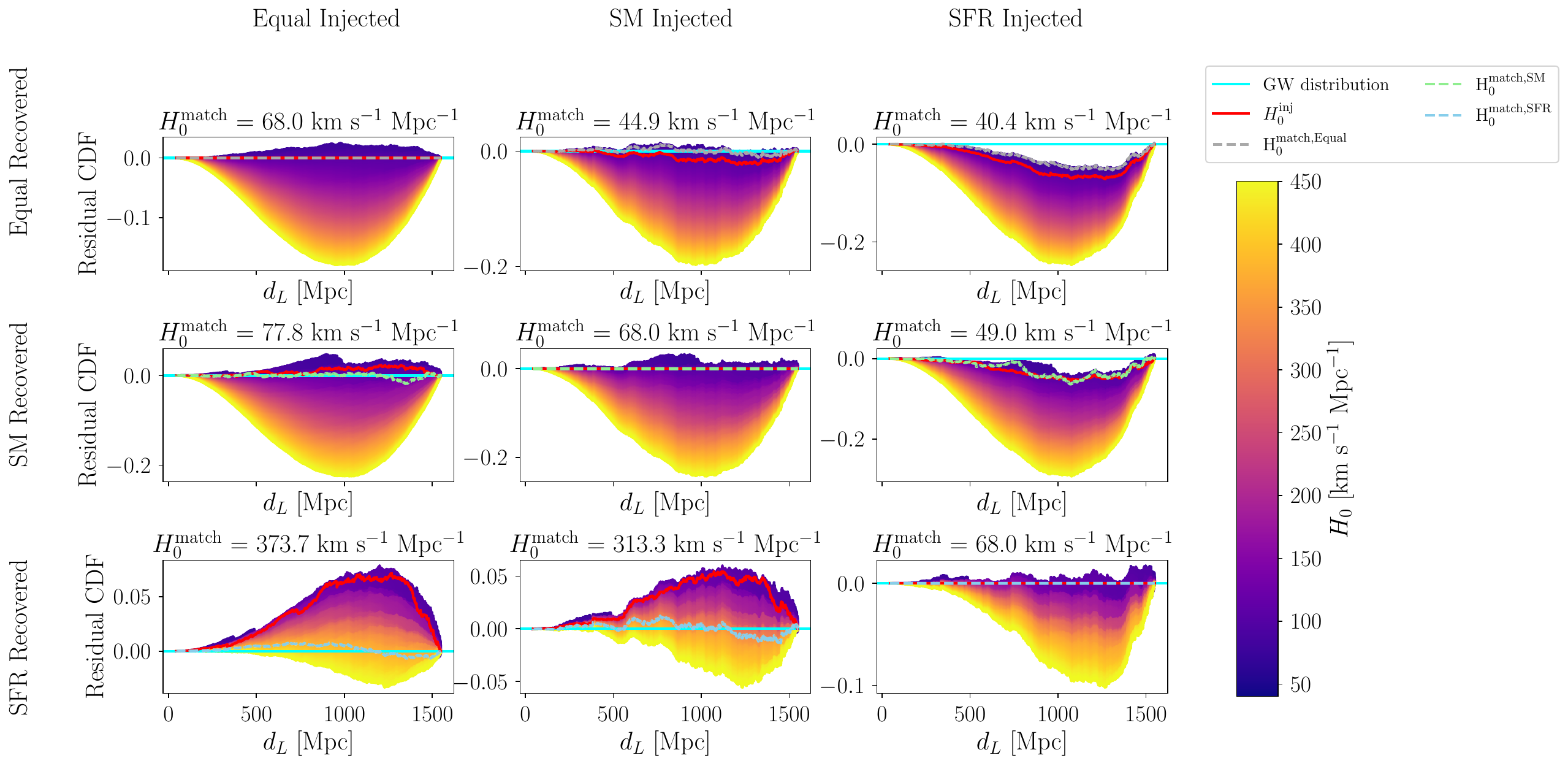}
\caption{Residual CDFs when comparing a measured GW luminosity distance distribution {(with $A=0$)} following a given true injected distribution (cyan; Residual CDF = 0) with an assumed weighted galaxy redshift distribution, converted to a luminosity distance distribution for a given value of $H_0$. The comparison when using the injected value of H$_0^{\rm inj}$ is shown in red for all injections and recoveries. The $H_0$ value that minimizes the residual, $H_0^{\rm match}$, for each combination is reported in the titles and denoted by the {gray, light blue, and light green} dashed lines {corresponding to a recovery weighting scheme following equal, SFR, or SM weights}.}
\label{fig:dist-matching}
\end{figure*}

We also investigate the influence of uncertainties by looking at cases where GW luminosity distance measurements have a 1$\%$, 10$\%$, or 20$\%$ error, as well as the influence of the number of galaxies in our lines of sight, $N_{\rm GAL}$. Fig.~\ref{fig:Bias-per-pcerr} shows the resulting $H_0$ posteriors for each injection-recovery set for different runs with different values of $N_{\rm GAL}$, for all fractional errors considered. We discern three different regimes that may lead to significant systematic biases in the inference. We denote these three regimes as the ``well-localized" regime occurring at low $N_{\rm GAL}$, the ``transitional" regime occurring at moderate $N_{\rm GAL}$, and the ``uninformative" regime occurring at large $N_{\rm GAL}$, and describe the results in Fig.~\ref{fig:Bias-per-pcerr}. In Appendix \ref{sec:mockcats}, we create four other mock catalogs, each removing one feature of \textsc{UniverseMachine} at a time: differing weighted galaxy redshift distributions, correlations between SM and SFR, low amounts of highly-weighted galaxies, and the volumetric effects of having a three-dimensional universe. We see how each of these features change the biases in these three regimes to identify which effects determine the biases we see. We describe in more detail these three identified regimes below, but leave further details on how these regimes were identified to Appendix \ref{sec:mockcats}.

\subsection{Uninformative regime}
When in a regime where the {number of galaxies along each line of sight} 
is high ($N_{\rm GAL} \gtrsim 10^5${; see Fig.~\ref{fig:Bias-per-pcerr}}), the posteriors in $H_0$ fall into the ``uninformative" regime. When in this regime, any information from any one galaxy is washed out, and all the information seen in the $H_0$ posteriors comes from matching the total observed GW luminosity distance distribution with the assumed weighted galaxy redshift distribution (see e.g. \citealt{2019JCAP...04..033D}, \citealt{2021PhRvD.104d3507Y}). To demonstrate this effect, we compute the residual (difference) between the GW observed luminosity distance cumulative distribution function (CDF{; $A=0$}) and the weighted galaxy luminosity distance distribution (converted from the redshift distribution for a range of $H_0$ values) CDF for luminosity distances between zero and $\hat{d_L}^{\rm thr}$. We plot these residuals in Fig.~\ref{fig:dist-matching}. When we assume the true (injected) weighting scheme, the GW luminosity distance distribution will follow the injected weighted galaxy redshift distribution, and we recover the correct $H_0$ value. However, when we weight the galaxy redshifts with the incorrect weighting scheme, the GW luminosity distance distribution will not match up with the weighted redshift distribution for the correct $H_0$ value (see the red lines in Fig.~\ref{fig:dist-matching}), but will match up for some different value of $H_0$ (see $H_0^{\rm match}$ and the {light blue, light green, and gray} dashed lines in Fig.~\ref{fig:dist-matching}). In Fig.~\ref{fig:Bias-per-pcerr}, for each injected distribution, we plot {these best `matched'} $H_0$ values for equal, SM, and SFR recovery weights as gray, light green, and light blue horizontal lines, respectively. As seen in Fig.~\ref{fig:Bias-per-pcerr}, as the number of galaxies {along each line of sight} increases, the $H_0$ posteriors for each recovery weighting scheme trend towards the horizontal lines indicating what value of $H_0$ matches the GW luminosity distance distribution to the weighted galaxy redshift distributions. Likewise, as is shown in Fig.~\ref{fig:Bias-per-pcerr}, for the same value of $N_{\rm GAL}$, increasing the fractional error of the GW luminosity distance measurements also follows the same trend.

\subsection{Well-localized regime}
Now we consider the opposite limit --- the ``well-localized" regime where $N_{\rm GAL}$ is small. Note that in the limit where {there is only} one galaxy along each line of sight ($N_{\rm GAL} = 1$), we de facto identify the host galaxy and therefore are in effect pursuing the bright siren method. {Even in the limit where $N_{\rm GAL} > 1$ but below a certain threshold ($N_{\rm GAL} \lesssim 100$), as is apparent in Fig.~\ref{fig:Bias-per-pcerr}, we see that there are no biases regardless of what weighting scheme is used during the inference. The reason for this depends on what the true (correct) weighting scheme is. First, let us consider the case where galaxies all have an equal probability of hosting GW events.} In this regime, we find that, on average, {there are very few lines of sight containing GW events that contain} any extremely highly weighted galaxies {(i.e. galaxies with large SM or SFR)}
such that, when combining posteriors, our final inference on $H_0$ is unbiased for all recovery weighting schemes, regardless of whether it is the correct scheme. 

This argument is especially visible in a mock catalog, \textsc{Uniform:UMuncorrelated}, that we introduce in Appendix \ref{sec:mockcats}. In this catalog, we {remove} volumetric effects and assign the `stellar mass' to be either one or a thousand with $1\%$ of galaxies having a weight of a thousand in order to clearly identify which galaxies are considered ``highly-weighted" and isolate the bias due only to these highly-weighted galaxies. 
In this case, each line of sight will have $\mathcal{O}(1)$ high-mass galaxy if $N_{\rm GAL}$ $\gtrsim$ 100. The ``well-localized" regime in this case is when $N_{\rm GAL}$ $\lesssim$ 100, where we see in Fig.~\ref{fig:mock-cat-results} that $N_{\rm GAL}$ $\lesssim$ 100 does recover an unbiased $H_0$ estimate for all injection-recovery combinations for the \textsc{Uniform:UMuncorrelated} mock catalog. 

This trend is more difficult to see in more realistic mock catalogs, such as our \textsc{UniverseMachine} catalog, where the galaxy properties have smooth one-dimensional distributions. However, we can provide a rough estimate of the ``well-localized" regime limit by the following logic:
let us assume that our mock catalog has $\epsilon \%$ of galaxies that have $x$ orders of magnitude higher weights (than the majority of galaxies). Then, we know that there will be on average no highly-weighted\footnote{{``Highly-weighted" in this case means a galaxy with a weight above $10^x$ times more than the majority of galaxies.}} galaxy along each line of sight if $N_{\rm GAL} \lesssim 100/\epsilon$. However, for any bias from a highly-weighted galaxy to be relevant, that galaxy would need to have sufficient weight to overwhelm the posterior support from the rest of the galaxies along the line of sight. We use this logic to combine the above condition with the condition that $N_{\rm GAL}$ $\lesssim 10^x$. We can illustrate this with a back-of-the-envelope calculation using the SM distribution in the \textsc{UniverseMachine} mock catalog. If we assume that the true distribution of GW events is equally likely to be in any host galaxy, the majority of {host} galaxies {will} have ${\rm SM } \sim 10^8 M_\odot$, with $\sim 0.55\%$ of {host} galaxies having ${\rm SM }\sim 10^{11} M_\odot$. However, if we weight following SM, the majority of {host} galaxies {will} have ${\rm SM }\sim 10^{11} M_\odot$ (see e.g. Fig.~\ref{fig:UMLCdists}). Then, with the above estimate, we would expect the ``well-localized" regime to end around $N_{\rm GAL} \sim \frac{100}{0.55} \sim 180$. As we see in left panel of Fig.~\ref{fig:Bias-per-pcerr}, this estimate is fairly accurate in predicting where the biases begin to appear.

{Let us now consider the case where host galaxies follow either SM or SFR weights. When generating GW events, most will have large SM (or SFR) values. Therefore, when using the lines of sight that contain GW events to find a posterior on $H_0$, each line of sight will have on the order of one large SM (or SFR) galaxy if $N_{\rm GAL} \lesssim 100/\epsilon$ as before, which is typically the true host. When recovering with any weighting scheme, since there are so few galaxies along each line of sight, the true host will always have non-negligible support. Thus, in this case, the usual argument of dark sirens apply and the true $H_0$ value will appear from combining many observations.}

\subsection{Transitional regime}
Finally, in between the ``uninformative" and ``well-localized" regimes, there is a third regime we have termed the ``transitional" regime. In this regime, {regardless of the true host galaxy weighting scheme}, any line of sight will have on the order of a few highly-weighted galaxies. The $N_{\rm GAL}$ range that determines this regime depends on the percentage of highly-weighted galaxies in our mock catalog as well as the  fractional error in the GW luminosity distance. In the case of our mock \textsc{UniverseMachine} catalog, this regime corresponds to $N_{\rm GAL} \sim 100\mbox{--}10,000$ when the GW luminosity distances have a $10\%$ error (see Fig.~\ref{fig:Bias-per-pcerr}). Along a given line of sight, these few extremely large galaxies have weights that are several orders of magnitude higher than other galaxies, and thus they dominate the posterior for that line of sight. If the GW distribution follows SM or SFR, the true host in any line of sight will likely be in one of these few highly weighted galaxies. If we recover with equal weights in this regime, there is still some support for the correct host along each line of sight and the usual dark siren argument holds such that we will still recover the correct $H_0$. On the other hand, if the true distribution is such that all galaxies in the catalog are equally likely to be hosts, recovering with SM or SFR leads to extremely large biases. {This effect when moving from the ``well-localized" to the ``transitional" regime can be seen by the very sharp jumps in $H_0$ bias in Fig.~\ref{fig:Bias-per-pcerr}, especially in the leftmost plot.} These potentially large biases appear when we consider that all galaxies along the line of sight contribute to the final $H_0$ posterior, regardless of whether the galaxy falls within the localization volume of the GW event (or below $\hat{d_L}^{\rm thr}$) for the injected $H_0$ value. This effect is easiest to see again in the \textsc{Uniform:UMuncorrelated} mock catalog. In this catalog, galaxies are distributed uniformly along a line in redshift. A $\hat{d_L}^{\rm thr}$ of 1550 Mpc corresponds to a redshift of $z \sim 0.3$. However, since our galaxy catalog extends to a redshift of 1.4, when highly-weighted galaxies are distributed along the line of sight, the majority of these galaxies will be at a redshift $z > 0.3$. When inferring $H_0$, these galaxies will give support for $H_0 > H_0^{\rm inj}$, leading to a bias to high $H_0$ values. While the \textsc{Uniform:UMuncorrelated} mock catalog does not contain any volumetric effects (this catalog only considers a one-dimensional line in distance), when considering more realistic mock catalogs such as our \textsc{UniverseMachine} mock catalog, this argument still holds.

\section{Discussion of potential biases} \label{sec:implications}
In \S\ref{sec:results} we found that assuming an incorrect galaxy host probability distribution can lead to large biases in the recovered value of $H_0$. In this section, we discuss potential factors that may influence these biases, as well as demonstrate a diagnostic that can be used to identify when an incorrect weighting scheme is used. 

\subsection{Improving GW localization} \label{sec:pc-err}
As seen in Fig.~\ref{fig:Bias-per-pcerr}, as we increase the  fractional error in GW luminosity distance, we approach the ``uninformative" regime. Likewise, this means that decreasing the fractional error will allow us to stay in the ``well-localized" regime for longer (i.e. to higher $N_{\rm GAL}$). In other words, if we have better GW localization, there is a better chance that we may be in the ``well-localized" regime and recover unbiased estimates of $H_0$. Intuitively, this is the same as keeping the same fractional error and decreasing the number of galaxies in the localization volume. Therefore, with future GW detectors (e.g. Cosmic Explorer \citep{2021arXiv210909882E} or Einstein Telescope \citep{2020JCAP...03..050M}), we may be able to mitigate possible biases if we only consider events with small GW luminosity distance uncertainties or sky localization such that we are in the ``well-localized" regime. However, implementing such a selection would give rise to a non-trivial $P^{\rm GW}_{\rm det}$, making it difficult (but not impossible) to correct for selection biases.

\begin{figure}[tbh]
\centering
\includegraphics[width = \linewidth]{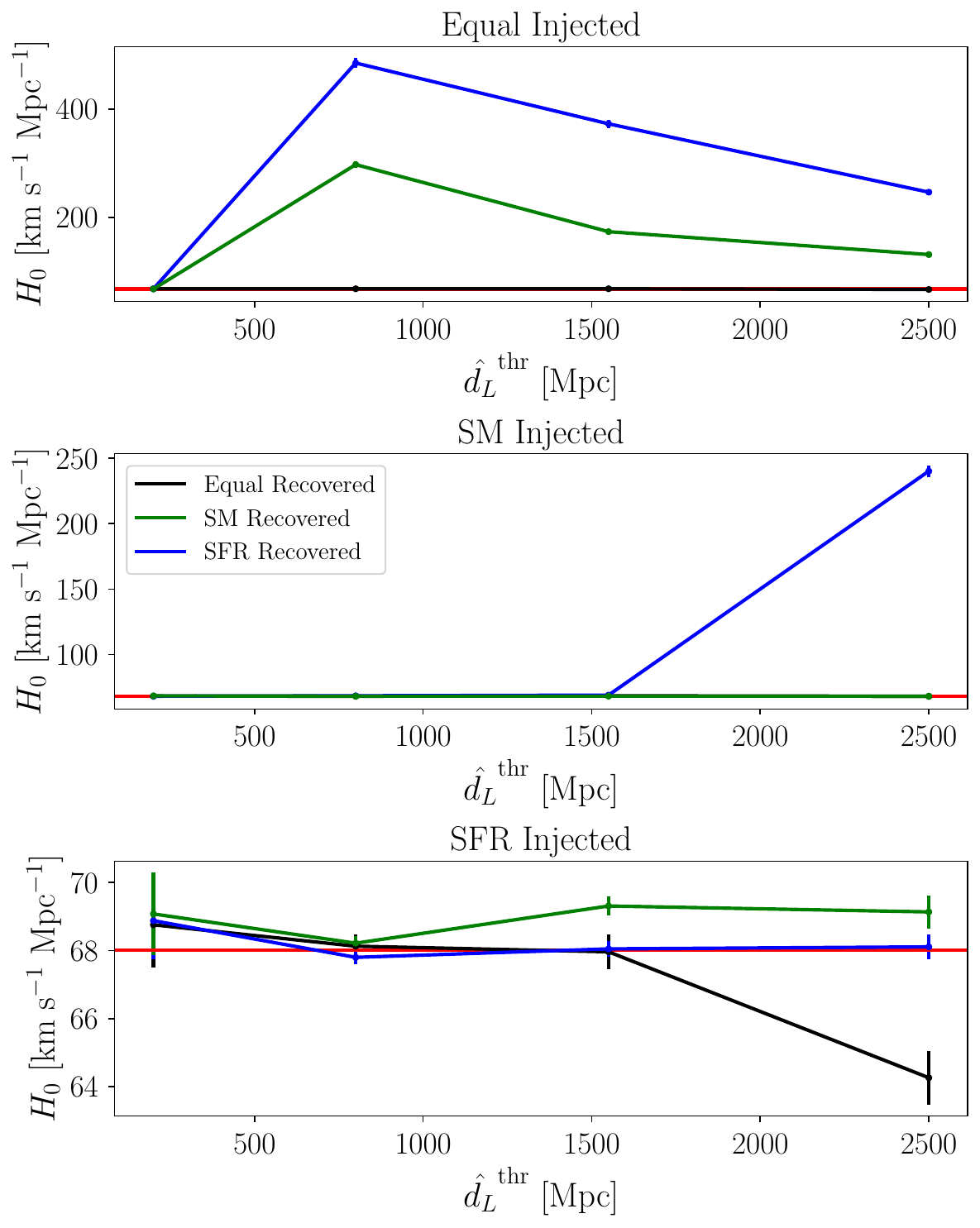}
\caption{Bias in $H_0$ inference when changing the GW detection threshold, $\hat{d_L}^{\rm thr}$. The posteriors were generated from 10,000 {GW events coming from lines of sight} containing 5,000 galaxies. {The} GW events {each have a} $10 \%$ fractional error {in luminosity distance}.}
\label{fig:dLthr-CDFs}
\end{figure}

\subsection{Exploiting correlations}\label{sec:correlationDiscussion}

As seen in Appendix \ref{sec:mockcats}, the correlation between SM and SFR helps mitigate the bias when injecting with one and recovering with the other. If there were to be no correlation between SM and SFR, as is the case in the \textsc{UM:UMuncorrelated} catalog in Appendix \ref{sec:mockcats}, we see that the biases in these two injection-recovery cases are much larger. Since there are correlations in parameter space in actual galaxy catalogs, it is possible that weights built using some optimal combination of galaxy properties could alleviate these biases. {For example, if we assume \textsc{UniverseMachine} correctly depicts our universe, using SM weights may be the best choice to minimize any biases, as long as true host galaxies do not have equal weights.} However, a deep knowledge of such a catalog would be needed to build a weighting scheme that exploits these correlations, which is not currently known with present catalogs.

\subsection{Higher SNR cut} \label{sec:SNR}
In our toy prescription, a lower GW detection threshold $\hat{d_L}^{\rm thr}$  corresponds to a higher SNR cut. One would expect that, due to GW luminosity distances having a constant fractional error, decreasing the detection threshold would lead to better localized events. As discussed in \S\ref{sec:pc-err}, better localization corresponds to staying in the ``well-localized" regime longer, but {can lead} to larger biases in the ``transitional" regime. In Fig.~\ref{fig:dLthr-CDFs}, we investigate the level of biases seen for different injection-recovery schemes when there are 5,000 galaxies along each line of sight. In general, we see in Fig.~\ref{fig:dLthr-CDFs} that sufficiently decreasing the GW detection threshold does help mitigate some biases, although this may not be the case in general (see for example in the top panel of Fig.~\ref{fig:dLthr-CDFs}; when we inject equal weights, decreasing the detection threshold from the original $1550 \, \rm Mpc$ increases the biases until the threshold reaches about $800 \, \rm Mpc$, where it then starts decreasing).

\subsection{Galaxy clustering}\label{sec:LSS}
We have not considered the effect of galaxy clustering in this analysis, although it might help mitigate the magnitude of the biases when using an incorrect weighting scheme. While we cannot make any claims on the complete effects of large-scale structure, the mitigating effects of parameter correlations (eg. SM and SFR correlations) seem to, at first order, help decrease the magnitude of biases we see. This supports the claim that further correlations may, again, decrease the biases. However, \cite{Perna2024} demonstrate using the MICECAT catalog---a mock catalog that accounts for galaxy clustering---that there are still biases in the inference of $H_0$ when assuming an incorrect weighting scheme when there are anisotropies in the galaxy structure. \cite{Gairetal2023} mention that, over different lines of sight, the over/underdensities in catalogs that contain anisotropic structure would average out. Thus, as seen in \cite{Perna2024}, there are still biases present due to mismatching the weighted galaxy redshift distributions. We also investigate potential biases using the MICECAT mock catalog using our scheme above {in Appendix~\ref{sec:MICECAT}} and find similar biases to our above results. On the other hand, it is possible that if different weighting schemes prefer the same structure, this positive correlation may lead to an unbaised estimate of $H_0$, and may even improve it.

\subsection{Diagnosing incorrect assumptions}\label{sec:hierarchicalAnalysis}
\begin{figure}
\centering
    \includegraphics[width = \linewidth]{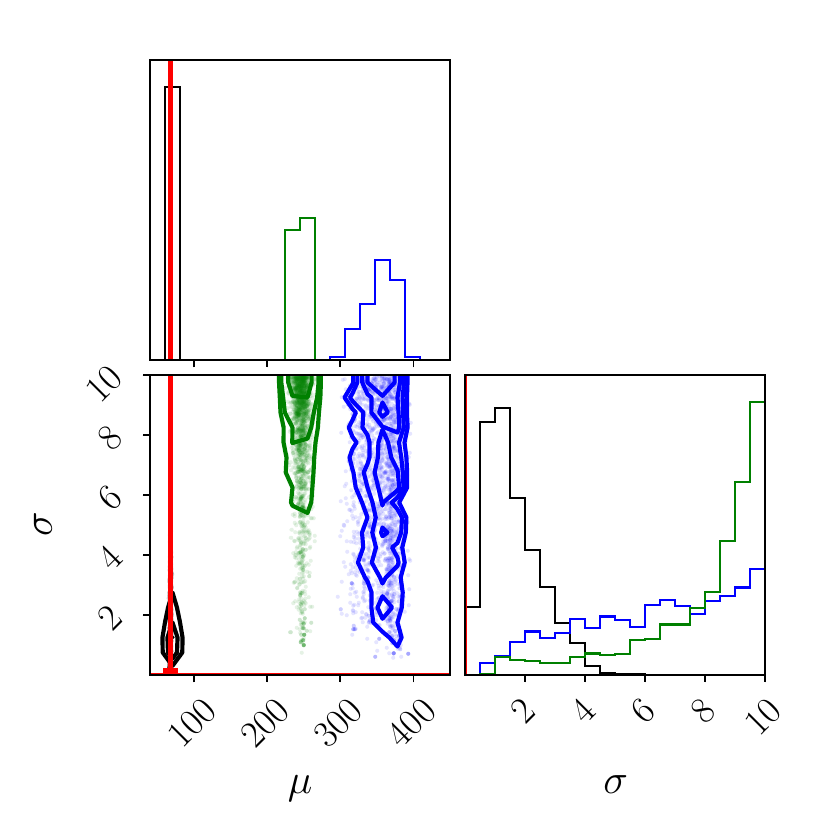}
\caption{Example posteriors on $H_0$ inference using hierarchical analysis assuming $H_0$ is a Gaussian with mean $\mu$ and standard deviation $\sigma$. The true distribution assumes that host galaxies are all equally likely to be hosts. Each line of sight contains 2,500 galaxies and we observe {2,500} GW events, each with 10\% error in luminosity distance. We see that recovering with SM or SFR causes the recovered standard deviation to shift away from zero, and the mean shifts away from the true value ($H_0^{\rm inj} = 68 \; {\rm km\, s^{-1} \, Mpc^{-1}}$). {This suggests that using the population of GW events can diagnose when an incorrect weighting scheme is used (and can help ``self-calibrate" the correct weighting scheme, thereby mitigating the bias due to incorrect galaxy weighting; see \S~\ref{sec:hierarchicalAnalysis} and~\cite{Vijaykumar:2023bgs})}.
}
\label{fig:diagnosis}
\end{figure}

We assume a single value of $H_0$, which allows us to use the methods described throughout this text in our inference. However, in the event that we use an incorrect galaxy host weighting scheme, our analysis need not find that a common value of $H_0$ is recovered by the population of GW events. As shown in \cite{2019PhRvD..99l4044Z}, multiplying the individual event likelihoods on $H_0$ in the latter case fails to capture the deviations present in the sample. Therefore, we suggest using hierarchical analysis~\citep[{such as in}][]{2019PhRvL.123l1101I} to recover $H_0$ in order to test for model misspecification such as an incorrect galaxy weighting scheme. {For instance, we could posit that instead of the true value of $H_0$ being a constant, the value of $H_0$ is drawn from a Gaussian distribution with mean $\mu$ and standard deviation $\sigma$.}
If we recover using the correct weighting scheme, we expect to recover a delta-function with $\sigma = 0$ and $\mu = H_0^{\rm inj}$. However, if we recover using the incorrect weighting scheme, we may see that our recovered $H_0$ value does not converge, or converges with a non-zero standard deviation. An example plot is shown in Fig.~\ref{fig:diagnosis} demonstrating this effect. We inject equal weights for the host galaxy probabilities, and find that recovering with equal weights indeed results in a 
posterior at $H_0^{\rm inj} = 68 \, {\rm km\, s^{-1} \, Mpc^{-1}}$ {with support for $\sigma = 0$}\footnote{{The posterior peaks slightly away from $\sigma=0$ while assuming the correct recovery weighting scheme. This is likely because we are trying to probe a very narrow feature (essentially a Dirac delta function) with hierarchical inference while using finite number of samples to construct the relevant Monte Carlo sums (see e.g.~\citealt{2022arXiv220400461E} for a description of this effect). Increasing the number of samples does shift the peak towards $\sigma=0$, but also increases the computational cost.}}. However, recovering with weights following either SM or SFR leads to posteriors that have non-zero standard deviations and peak away from the injected $H_0$ value. While this method is extremely useful in diagnosing if an incorrect weighting scheme is used, we emphasize that using hierarchical analysis does not diminish the bias, nor does it give any information on what the correct weighting scheme should be. {However,} it should be possible to simultaneously infer the weighting scheme as well as $H_0$ by generalizing the idea laid out in \cite{Vijaykumar:2023bgs}; however, we do not investigate this here.

\subsection{Decreasing the $H_0$ prior}\label{sec:small-prior}
We emphasize that our above analysis uses a conservative prior of $H_0 = [40 ,450] \, {\rm km\, s^{-1} \, Mpc^{-1}}$. The current landscape of Hubble constant measurements---e.g. from measurements of the CMB~\citep{2020A&A...641A...6P}, supernovae~\citep{2022ApJ...938..113S}, or the tip of the red giant branch~\citep{2020ApJ...891...57F}---reliably place measurements of $H_0$ to be within $\sim 60\mbox{--}80 \, {\rm km\, s^{-1} \, Mpc^{-1}}$. While we see that restricting our $H_0$ prior to these values in the current analysis does not change the biases we see (the posteriors simply rail against the prior bounds), if we use a smaller prior along with the hierarchical inference of $H_0$ suggested in \S\ref{sec:hierarchicalAnalysis}, the correct $H_0$ value may be inferred, albeit with some non-zero standard deviation{, due to the restrictive prior absorbing some of the systematic uncertainty present when using an incorrect weighting scheme.} 
{While using hierarchical analysis with a very restrictive $H_0$ prior may yield unbiased results, }
it may fail if individual GW event posteriors are largely uninformative over this smaller prior range. In that case, the hierarchical analysis would infer a mean at the center of the prior range and a large standard deviation that encompasses the entire prior. {Therefore, while we recommend always using hierarchical analysis in future inferences due to its strength in diagnosing incorrect weighting schemes, incorporating a restrictive $H_0$ prior may additionally yield unbiased results regardless of weighting scheme. However, we emphasize that additional investigations need to be carried out to confirm if this is always true.}

\section{Conclusion} \label{sec:discussion}
We have examined the dark siren approach to cosmology, wherein all galaxies in a binary's localization volume are considered as potential hosts to a given source. In particular, we have explored the impacts of weighting the galaxy catalog incorrectly, and find the potential for substantial biases in the inferred value of $H_0$. We break these biases into three regimes determined by the {number of galaxies in each line of sight, }
$N_{\rm GAL}$: the ``well-localized," ``transitional," and ``uninformative" regimes. We  create multiple galaxy catalog toy models to isolate each effect that might influence the observed biases. We advocate the use of hierarchical analysis during $H_0$ inference to help diagnose any potential biases, as any noticeable standard deviation in the $H_0$ posterior identifies the use of an incorrect weighting scheme. We find that correlations between parameters such as stellar mass and star formation rate, as well as correlations with large-scale structure, may reduce potential biases. We also note that future GW detectors that improve GW luminosity distance localization may help mitigate some of these biases, so long as the number of galaxies along any given line of sight remains small. Note that our results also assume a $100\%$ complete galaxy survey; a realistic survey will be incomplete, and the choice of weights should also be taken into account while correcting for catalog incompleteness.

Finally, we note that current LVK analyses \citep{2023ApJ...949...76A} do not find any substantial bias in the recovered $H_0$ value compared to conventional, non-standard siren  determinations. There are a number of reasons for this unbiased determination. First, as reported by~\cite{2023ApJ...949...76A}, current constraints of $H_0$ from galaxy catalogs with K-band weighting give a $\sim18\%$ measurement, while a spectral siren analysis using a fixed GW population without galaxy catalogs yields a $\sim20 \%$ measurement of $H_0$. This demonstrates that LVK analyses using the dark siren method are presently dominated by uncertainties associated with the GW population, and hence, any incorrect weighting scheme when using galaxy catalogs would be expected to have a subdominant effect. In addition, the current luminosity weighting schemes may be well-informed (see e.g.  \cite{Vijaykumar:2023bgs}, which constrains host galaxies from the evolution of the GW merger rates), such that current LVK analyses may be using a sufficiently accurate weighting such that the bias is minimized. \corrections{Additionally, a joint inference of the binary population properties and $H_0$~\citep{2023JCAP...12..023G, 2023PhRvD.108d2002M} could help mitigate this bias, although consistency should be ensured between the galaxy weights and the redshift evolution of the merger rate~\citep{Vijaykumar:2023bgs} while constructing the likelihood function.} Finally, large-scale clustering in the galaxy catalogs currently in use by the LVK may also help to mitigate these biases. Even so, as our data improves, the bias in $H_0$ due to incorrect galaxy weighting may become an increasing concern for dark siren approaches. Our work highlights the importance of accounting for and mitigating the bias due to incorrect galaxy weighting in future dark siren measurements.

\section{acknowledgments}
{We thank the anonymous referee for a very careful review of our manuscript.} We also thank Gabriele Perna for helpful suggestions. AGH is supported by NSF grants AST-2006645 and PHY2110507, and gratefully acknowledges the ARCS Foundation Scholar Award through the ARCS Foundation, Illinois Chapter with support from the Brinson Foundation.
AV and MF acknowledge support to CITA by the Natural Sciences and Engineering Research Council of Canada (NSERC) (funding reference number 568580). AV is also supported by a Fulbright Program grant under the Fulbright-Nehru Doctoral Research Fellowship, sponsored by the Bureau of Educational and Cultural Affairs of the United States Department of State and administered by the Institute of International Education and the United States-India Educational Foundation.  
DEH is supported by NSF grants AST-2006645 and PHY-2110507, as well as by the Kavli Institute for Cosmological Physics through an endowment from the Kavli Foundation and its founder Fred Kavli.
This work has made use of CosmoHub~\citep{2020A&C....3200391T, 2017ehep.confE.488C}.

\software{numpy~\citep{2020Natur.585..357H}, scipy~\citep{2020NatMe..17..261V}, matplotlib~\citep{2007CSE.....9...90H}, {astropy~\citep{2013A&A...558A..33A,2018AJ....156..123A,2022ApJ...935..167A}}, jupyter~\citep{2016ppap.book...87K}, pandas~\citep{mckinney-proc-scipy-2010}, jax~\citep{jax2018github}, numpyro~\citep{phan2019numpyro,bingham2019pyro}.}
\clearpage
\appendix

\section{{Example of bias using the MICECAT mock galaxy catalog}}\label{sec:MICECAT}

\begin{figure}[bht]
    \centering
    \includegraphics[width = \linewidth]{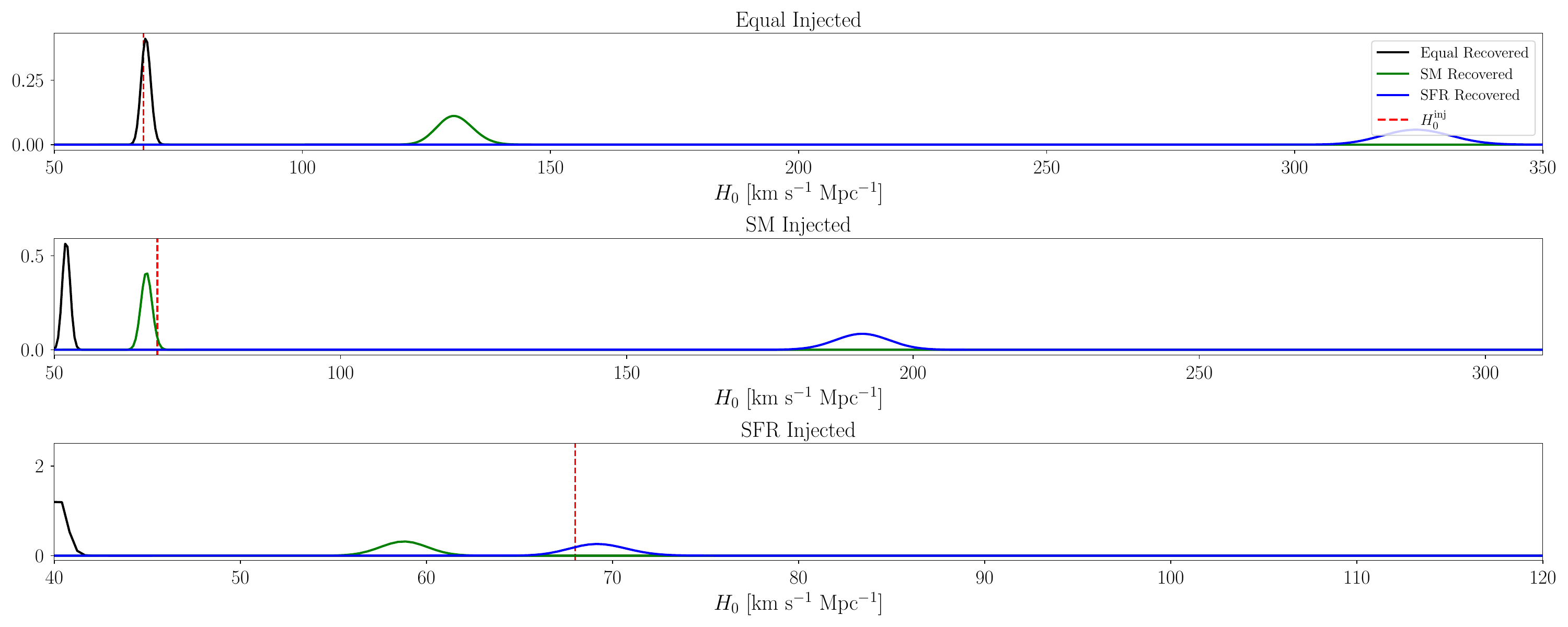}
    \caption{{Inference in $H_0$ using the MICECAT mock catalog from 20,000 GW events, with each line of sight containing 10,000 galaxies. Each GW event has a $20\%$ error ($A=0.2$) in luminosity distance. Assuming the correct weighting scheme leads to an unbiased $H_0$ recovery, but assuming an incorrect weighting scheme can lead to extremely large biases. The magnitude of the biases seen here are similar to those seen when using the \textsc{UniverseMachine} mock catalog (see Fig.~\ref{fig:Bias-single-param}).}}
    \label{fig:MICECATposts}
\end{figure}

{To investigate the effects of galaxy clustering on potential biases in $H_0$ inference if the incorrect host galaxy weighting scheme is used, we use the same prescription as outlined in \S~\ref{sec:UMcat}, now using the second version of the MICECAT mock galaxy catalog.}
{Fig.~\ref{fig:MICECATposts} demonstrates the potential bias in $H_0$ when using the MICECAT catalog. In this analysis, each line of sight contains 10,000 galaxies, with 20,000 GW events observed, each with a fractional error of 20$\%$. These are the same parameters as used for the \textsc{UniverseMachine} analysis in Fig.~\ref{fig:Bias-single-param}. When comparing with Fig.~\ref{fig:Bias-single-param}, we see that using the MICECAT catalog yields similar biases as when using the \textsc{UniverseMachine} mock catalog. One difference is that, due to the greater discrepancy in the MICECAT catalog between the equally-weighted galaxy redshift distribution with the stellar mass-weighted and star formation rate-weighted redshift distributions, injecting with either SM or SFR weights and recovering with equal weights leads to a larger bias than is seen in Fig.~\ref{fig:Bias-single-param}. Another difference is that, while the magnitude of the bias in the SM injected---SFR recovered case is slightly less severe, the SFR injected---SM recovered bias becomes more severe. We note that this is only a proof of concept to demonstrate that biases using the MICECAT catalog are similar to those in the \textsc{UniverseMachine} catalog and not a full analysis of how galaxy clustering affects the results presented above.}


\section{Discovering trends in the systematic biases in $H_0$ inference using simple mock catalogs} \label{sec:mockcats}

\begin{figure}[tbh]
    \centering
    \includegraphics[width = \linewidth]{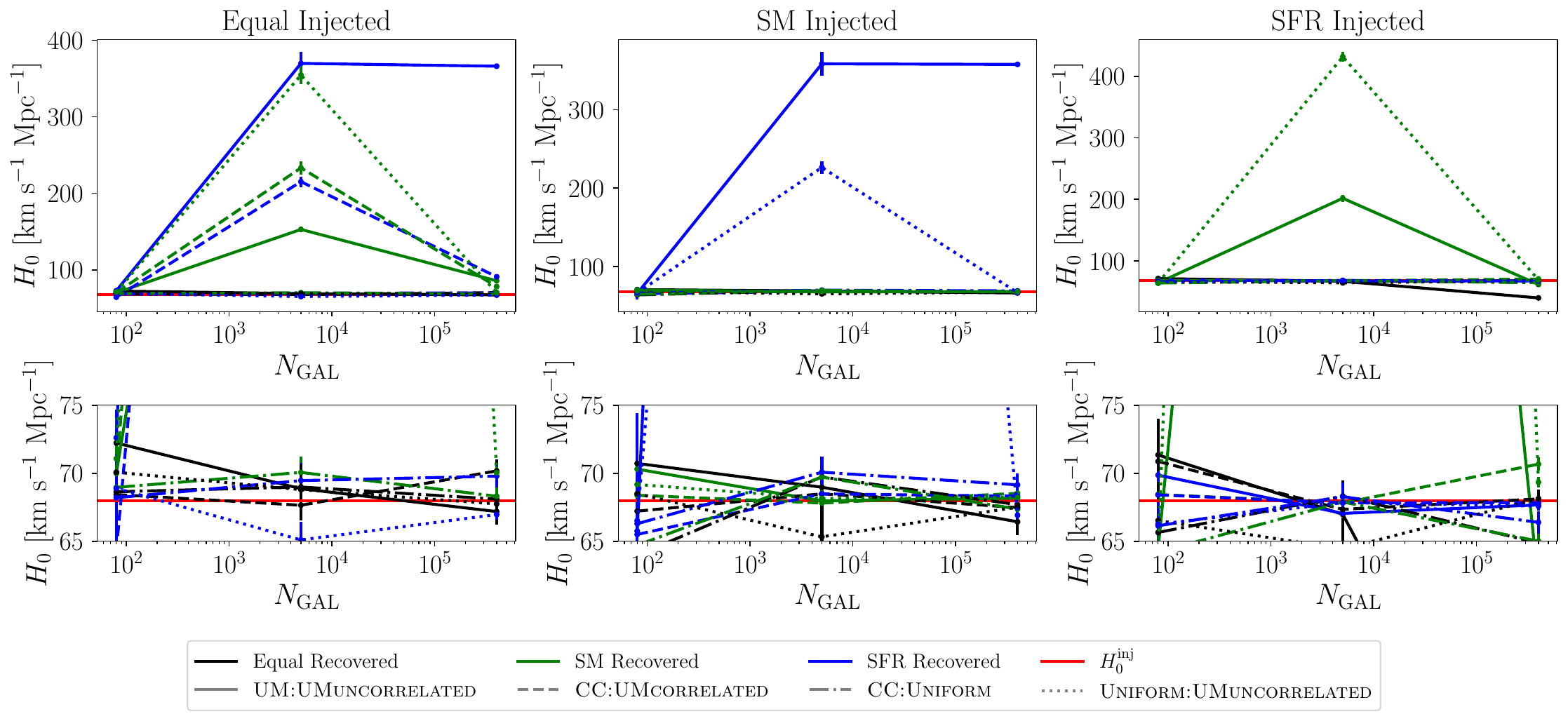}
    \caption{Inference on $H_0$ for four toy mock galaxy catalogs, \textsc{UM:UMuncorrelated} (solid), \textsc{CC:UMcorrelated} (dashed), \textsc{CC:Uniform} (dashdot), and \textsc{Uniform:UMuncorrelated} (dotted), for three $N_{\rm GAL}$, representing the ``well-localized," ``transitional," and ``uninformative" regimes. Recovering with equal weights is given by the black lines, recovering with stellar mass is given by the green lines, and recovering with star formation rate is given by the blue lines. {All GW events have a $10\%$ fractional error in luminosity distance. The errorbars correspond to the $1\sigma$ uncertainty in the posteriors when observing $N_{\rm GAL}/4$ GW events.}}
    \label{fig:mock-cat-results}
\end{figure}

To isolate different effects on the systematic biases seen in \S\ref{sec:results}, we create four more mock catalogs which decrease in levels of complexity. The catalogs are created by drawing a million galaxies from various redshift distributions for redshifts less than 1.4. However, the four catalogs differ in a couple of ways and are defined as follows:
\begin{enumerate}
\item \textsc{UM:UMuncorrelated} -- this catalog is created in the same way as is described in \S\ref{sec:UMcat}, but now SM and SFR are sampled from their marginalized one-dimensional distributions given by \textsc{UniverseMachine} such that there is no longer a correlation between SM and SFR, but each have the same weighted redshift distributions as in the original catalog.

\item \textsc{CC:UMcorrelated} -- Galaxies are drawn from a constant in comoving volume distribution, while stellar masses and star formation rates are assigned to each galaxy following the two-dimensional SM and SFR distribution given by \textsc{UniverseMachine}, such that SM and SFR are correlated with each other, but not correlated in redshift.

\item \textsc{CC:Uniform} -- galaxies are drawn from a constant in comoving volume distribution, and stellar masses and star formation rates are assigned independently to each galaxy, both following uniform distributions between the minimum and maximum value given by the \textsc{UniverseMachine} SM and SFR distributions. 

\item \textsc{Uniform:UMuncorrelated} -- this mock catalog ignores volumetric effects by distributing galaxies uniform in redshift in a one-dimensional Euclidean universe. Galaxies are assigned an SFR drawn from the marginalized one-dimensional SFR distribution given by \textsc{UniverseMachine}, while SM is assigned a weight of one or a thousand depending on a given percentage of highly-weighted galaxies, $\epsilon = 1 \%$.
\end{enumerate}
These mock catalogs are specifically chosen to investigate potentially significant effects independently from the full \textsc{UniverseMachine} mock catalog, such as influences from the correlation between SM and SFR (\textsc{UM:UMuncorrelated}), correlations of SM and SFR with redshift leading to different weighted redshift distributions (\textsc{CC:UMcorrelated}), and low-number statistics of highly-weighted galaxies (\textsc{CC:Uniform}). We also investigate any potential volumetric effects by creating a mock catalog that is set in a one-dimensional Euclidean universe (\textsc{Uniform:UMuncorrelated}).

Fig.~\ref{fig:mock-cat-results} shows posteriors on $H_0$ for all injection-recovery weighting schemes for three representative {total number of galaxies along each line of sight}, $N_{\rm GAL}$ = $\left[80,5000,400000\right]$, indicating the ``well-localized," ``transitional", and ``uninformative" regimes, respectively. The first {aspect} to note is that, even in the case of a one-dimensional Euclidean universe, recovering with the incorrect weighting scheme using the \textsc{Uniform:UMuncorrelated} catalog yields very large biases in the ``transitional" regime, but not in the ``well-localized" and ``uninformative" regimes. In the ``well-localized" regime {when the true hosts are all equally weighted}, on average, most of the lines of sight will not have a highly-weighted galaxy when recovering with SM or SFR due to the small $\epsilon$ percentage of large galaxies. In the ``uninformative" regime, the weighted galaxy redshift distributions are all equivalent, and thus we don't expect any bias in this case either. However, in the ``transitional" regime, since the galaxies are distributed uniformly in redshift, with a $\hat{d_L}^{\rm thr} = 1550 $Mpc corresponding to a $z \sim 0.3$, most of the highly-weighted galaxies will fall above this threshold for the injected $H_0$ value, but will be within the GW localization volume for a large $H_0$, leading to a bias to high $H_0$ as we see in Fig.~\ref{fig:mock-cat-results}. We do see that, for this catalog, recovering with equal weights when the true distribution follows SM or SFR in the ``transitional" regime remains unbiased.
This is because the redshift distributions are the same, and since all the weights are equal for all galaxies along the line of sight, recovering with equal weights will still have support for the correct host galaxy, albeit with less probability due to considering all galaxies along the line of sight.

On the other hand, we also see that all recovery weighting schemes using the \textsc{CC:Uniform} remain unbiased regardless of the {number of galaxies in each line of sight}. This is due to this catalog's SM and SFR distributions having a very large number of highly-weighted galaxies such that we are never in the ``transitional" regime. Since the weighted galaxy redshift distributions are all equivalent, we again see no biases in the ``uninformative" regime. However, when we consider more realistic stellar mass and star formation rate distributions, as in the \textsc{CC:UMcorrelated} catalog, the biases in the ``transitional" regime reemerge. In this catalog, SM and SFR are also correlated with each other, and we see that in this case, injecting with one and recovering with the other does not lead to any substantial biases.

The final catalog we consider is the \textsc{UM:UMuncorrelated} catalog, which contains the same weighted redshift distributions as the original \textsc{UniverseMachine} mock catalog we consider in the main text, but now SM and SFR are no longer correlated with each other. As we see in Fig.~\ref{fig:mock-cat-results}, the biases seen in all regimes matches those seen using the \textsc{UniverseMachine} mock catalog, but new biases emerge when injecting with either SM or SFR and recovering with the other. This effect demonstrates that correlations between SM and SFR can help mitigate the biases seen in the ``transitional" regime.

\bibliography{references}{}
\bibliographystyle{aasjournal}

\end{document}